\documentclass[aps, pre, twocolumn]{revtex4-1}
\usepackage{amssymb, amsfonts, amsmath, amsthm, mathtools, mathrsfs}
\usepackage{txfonts, pxfonts, upgreek}
\usepackage[usenames, dvipsnames]{xcolor}
\usepackage{braket}
\usepackage[colorlinks, linkcolor=blue, anchorcolor=red, citecolor=blue]{hyperref}
\usepackage{natbib}
\usepackage{array}
\usepackage{bigstrut}
\usepackage{inputenc}
\usepackage[T1]{fontenc}
\usepackage{lmodern}
\usepackage[english]{babel}

\begin{document}

\title{Full Counting Statistics and Fluctuation Theorem for the Currents in the Discrete Model of Feynman's Ratchet}

\author{Yu-Xin Wu\textsuperscript{1}}
\author{Jiayin Gu\textsuperscript{1}}\email{gujiayin@pku.edu.cn}
\author{H. T. Quan\textsuperscript{1,2,3}}\email{htquan@pku.edu.cn}

\affiliation{\textsuperscript{\rm 1}School of Physics, Peking University, Beijing 100871, China;}
\affiliation{\textsuperscript{\rm 2}Collaborative Innovation Center of Quantum Matter, Beijing 100871, China;}
\affiliation{\textsuperscript{\rm 3}Frontiers Science Center for Nano-Optoelectronics, Peking University, Beijing 100871, China}

\begin{abstract}
We provide a detailed investigation on the fluctuations of the currents in the discrete model of Feynman's ratchet proposed
 by Jarzynski and Mazonka in 1999. Two macroscopic currents are identified, with the corresponding affinities determined using 
 Schnakenberg's graph analysis. We also investigate full counting statistics of the two currents and show that fluctuation theorem 
 holds for their joint probability distribution. Moreover, fluctuation-dissipation relation, Onsager reciprocal relation and their nonlinear generalizations are numerically shown to be satisfied in this model.
\end{abstract}

\maketitle

\section{Introduction}

\par
\begin{figure}
\begin{minipage}[t]{0.8\hsize}
\resizebox{1.0\hsize}{!}{\includegraphics{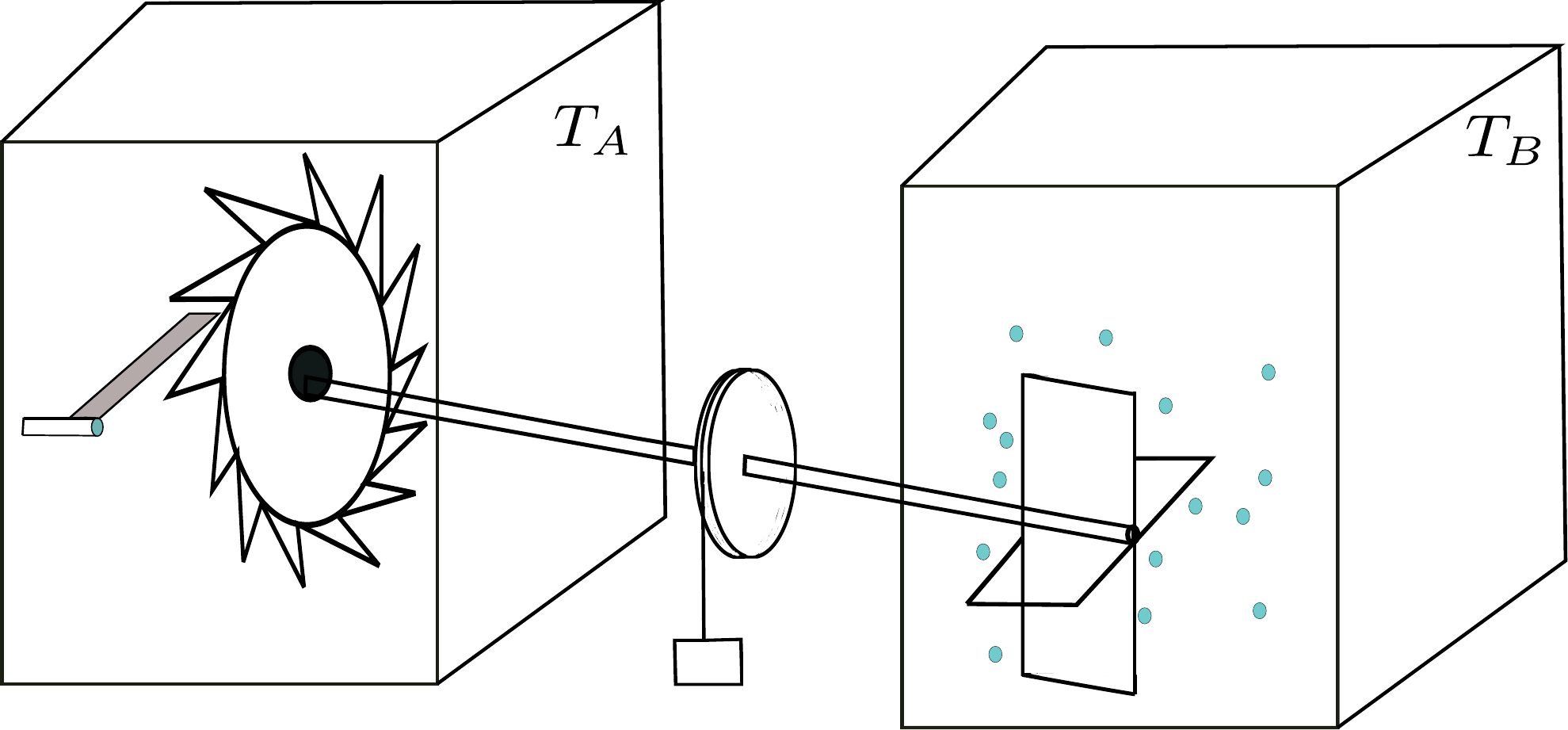}}
\end{minipage}
\caption{Schematic illustration of the Feynman's ratchet.}
\label{fig_ratchet}
\end{figure}

 Feynman's ratchet and pawl system -- a device capable of converting thermal random motion into useful work -- acts as
 a microscopic mechanical Maxwell's demon~\cite{Feynman_2010b}. As a paradigm for rectifying thermal noise, it has inspired extensive theoretical
 studies~\cite{Magnasco_PhysRevLett_1993,Parrondo_AmJPhys_1996,Sekimoto_JPhysSocJpn_1997,Magnasco_JStatPhys_1998,Astumian_EurBioPhysJ_1998,
 Broeck_PhysRevLett_2004,Ryabov_JStatPhys_2016,GomezMarin_PhysRevE_2006,Nakagawa_EurPhysLett_2006,Tu_JPhysA_2008,
 Hondou_JPhysSocJpn_1998,Komatsu_PhysRevE_2006,Skordos_AmerJPhys_1992} as well as experimental
 efforts~\cite{Martinez_SoftMatt_2017,Albay_OptExp_2018,Albay_ApplPhysLett_2020,
 Eshuis_PhysRevLett_2010,YongGunJun_PhysRevLett_2014,Berut_Nature_2012,Toyabe_JPhysSocJpn_2015}.
  Its basic setup includes a ratchet and a pawl (see Fig.~\ref{fig_ratchet}). From time to time, the pawl switches between being engaged and disengaged. The ratchet is attached
  to a windmill immersed in a heat reservoir at temperature $T_B$ where the gas molecules collide against the vanes of the windmill, causing the
   ratchet to rotate. The thermal motion of the microscopic pawl, which is immersed in another reservoir at temperature $T_A$, will cause the pawl
    itself to rise up or fall down from time to time. When it rises up, the pawl and the ratchet become disengaged, allowing 
    the free rotation of the ratchet in both directions. But when the pawl falls down, it will press against the teeth of 
    the ratchet, enforcing the ratchet to move in one direction only but never in the other. Feynman found the rates of rotation
     in two directions to be equal when the temperatures of the two reservoirs are equal. This is in accordance with the second
      law of thermodynamics. But when there is temperature difference between the two, the device can work as a heat engine or
       a refrigerator. In the heat engine "phase" the ratchet can rectify the thermal random motion of gas molecules to do useful work while
       in the refrigerator "phase" it can pump heat from the cold reservoir to the hot reservoir at the expenditure of work.

\par In 1999, Jarzynski and Mazonka proposed a very elegant discrete model that precisely captures the essential features 
of Feynman's ratchet~\cite{Jarzynski_PhysRevE_1999}, and it was recently realized experimentally~\cite{Bang_NewJPhys_2018}.
 Their model consists of only six states whose underlying dynamics is 
Markovian jump process described by a master equation. They calculated the mean values of mass velocity, energy flow and entropy
 production rate in nonequilibrium steady state. They also discussed the linear response relations for two currents--mass 
 displacement and energy flow. However, they didn't explore the fluctuation properties of the currents
 (but see Refs.~\cite{Sakaguchi_JPhysSocJpn_2000, Roeck_PhysRevE_2007}). As we know, Feynman proposed this automatic 
 mechanical version of Maxwell's demon to demonstrate that on average the second law can never be violated, but occationally, in a individual realization, 
 the second law can be probablistically "violated". Since the original Feynman's setting is too complex to be analyzed from the first principle,
  he himself was unable to quantitatively analyze the probability of "violating" the 
 second law.
 \par Over the last three decades, fluctuation theorems for systems in nonequilibrium steady state have been studied extensively,
   both theoretically~\cite{Nicolis_ProcNatAcadSciUSA_1971, Prigogine_1985, Evans_PhysRevLett_1993, Gallavotti_PhysRevLett_1995,
    Gallavotti_JStatPhys_1995, Kurchan_JPhysA_1998, Lebowitz_JStatPhys_1999, Maes_JStatPhys_1999, Jarzynski_PhysRevLett_1997,
    Jarzynski_PhysRevE_1997, Crooks_JStatPhys_1998} and experimentally~\cite{Wang_PhysRevLett_2002, 
    Seitaridou_JPhysChemB_2007, Utsumi_PhysRevB_2010, Kung_PhysRevX_2012, Garnier_PhysRevE_2005, Ciliberto_PhysRevLett_2013}. 
    In particular, Gaspard, Andrieux, and coworkers extended the fluctuation theorem to coupled currents~\cite{Andrieux_JChemPhys_2004,
     Andrieux_JStatMech_2006, Andrieux_JStatPhys_2007, Andrieux_NewJPhys_2009}. They provided 
       a unified framework for deducing fluctuation-dissipation theorem, Onsager reciprocal relations, and relations for
        higher order response coefficients from fluctuation theorem~\cite{Andrieux_JChemPhys_2004,Andrieux_JStatMech_2007,
        Gaspard_NewJPhys_2013,Barbier_JPhysA_2018}. Motivated by these results, it is desirable to explore the fluctuation
         properties of the Feynman's ratchet in the light of fluctuation theorems.
\par In this article, we focus on the fluctuation of currents, with the purpose of showing fluctuation theorem and its implications for 
the response properties of Feynman's ratchet model. By utilizing the method of graph decomposition proposed by Schnakenberg~\cite{Schnakenberg_RevModPhys_1976},
we decompose the currents in Feynman's ratchet model and calculate the cumulant generating function of the currents. With these results, we find the 
Gallavotti-Cohen symmetry of the currents. Thus we are able to quantitatively characterize the fluctuations and the probablistic "violation" of the second law
in Feynman's ratchet model. Our results confirm what Feynman has envisioned and deepen our understanding about the fluatuating properties of this
famous model from the qualitative way to the quantitative way.
\par The rest of the paper is organized as follows. Sec. II gives a detailed description of the discrete Feynman's ratchet
 model. Sec. III concerns the currents and the corresponding affinities. Sec. IV is devoted to the full counting statistics 
 of the currents. The response relations are numerically verified in Sec V. Conclusions are given in Sec. VI.

\section{Discrete Model of Feynman's Ratchet}

\par
\begin{figure}
\begin{minipage}[t]{0.8\hsize}
\resizebox{1.0\hsize}{!}{\includegraphics{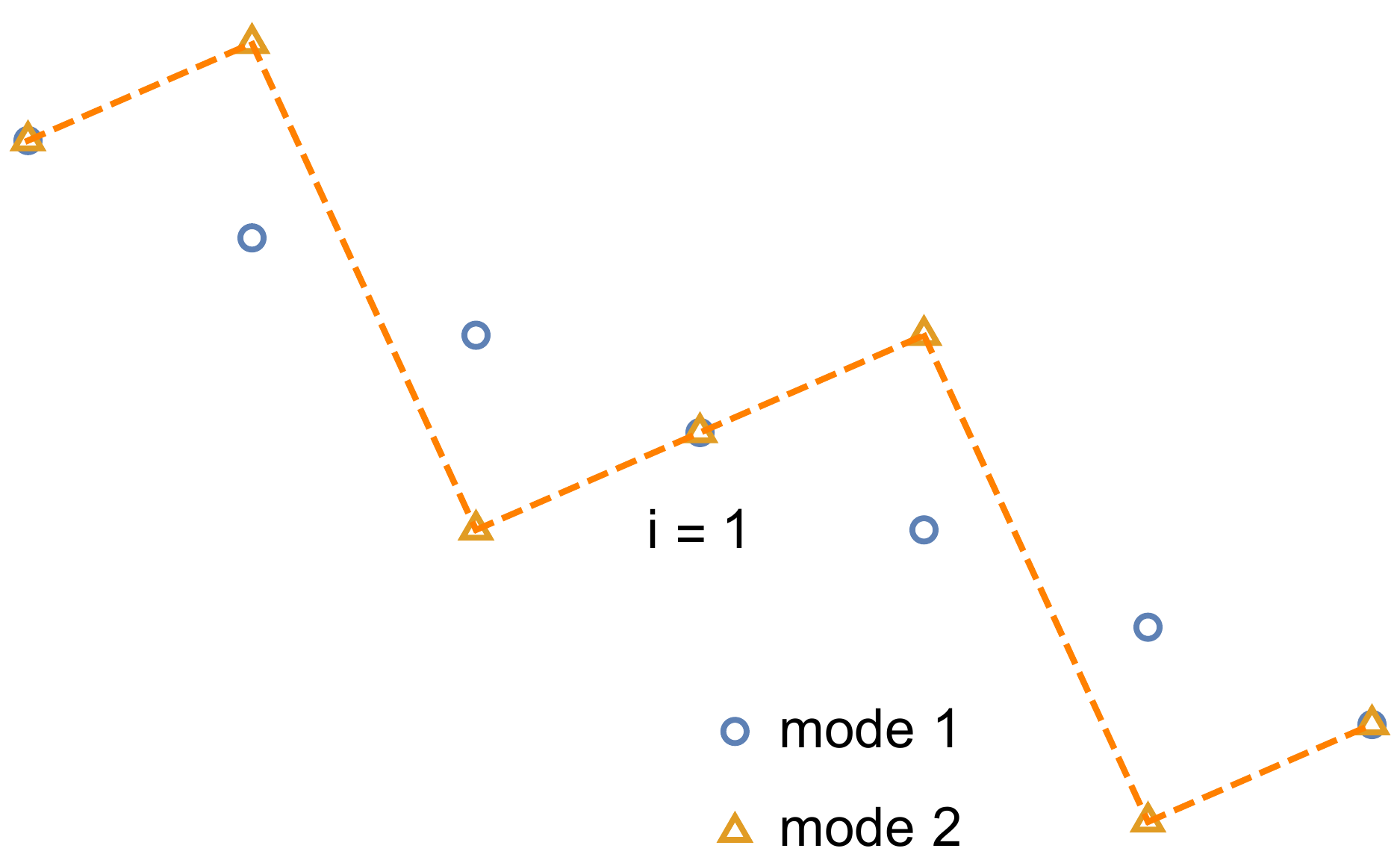}}
\end{minipage}
\caption{The landscape of potential energy $U_i^{(m)}$ for two modes, $m=1,2$, in the presence of an external load $f=\alpha/2d$.
Alternatively, we denote the potential energy as $U_{n}, n=1,\cdots,6.$ See Table~\ref{tab_E} for one to one correspondance. }
\label{fig_E}
\end{figure}

 The discrete model of Feynman's ratchet proposed by Jarzynski and Mazonka can be envisaged as a particle hopping between
 neighboring sites on a 1D regular lattice. As shown in Fig.~\ref{fig_E}, the potential energy of each site $i$ has two modes 
 $m=1,2$.
\begin{align}
U_i^{(m)}=\begin{cases}
-ifd \text{,} & m=1\text{,} \\
\alpha\cdot\left[(i\;\text{mod}\;3)-1\right]-ifd \text{,} & m=2 \text{.}
\end{cases}
\end{align}
Here, $f$ denotes the load, $\alpha$ is the energy unit and $d$ is the lattice spacing. Due to periodicity, we only need to 
consider one tooth on the ratchet: the system consists of six possible states, $n=1,\cdots,6$. The energies are listed in
 Table~\ref{tab_E}. Only those transitions between neighboring sites in the same mode and those between modes on the same site 
 are allowed.

\begin{table}
\caption{The energies of the six states.}
\begin{center}
\begin{tabular}{>{\centering\arraybackslash}m{1.5cm}|>{\centering\arraybackslash}m{1.5cm}|>{\centering\arraybackslash}m{1.5cm}|>{\centering\arraybackslash}m{3.0cm}}
\hline
\hline
state $n$   &  mode $m$     &  site $i$     &  potential energy $U_{n}$  \bigstrut \\ \hline
 1  &    1    &    0    &   $0$  \bigstrut \\ \hline
 2  &    1    &    1    &   $-fd$  \bigstrut \\ \hline
 3  &    1    &    2    &   $-2fd$  \bigstrut \\ \hline
 4  &    2    &    0    &   $-\alpha$  \bigstrut \\ \hline
 5  &    2    &    1    &   $-fd$  \bigstrut \\ \hline
 6  &    2    &    2    &   $\alpha-2fd$  \bigstrut \\ \hline
\hline
\end{tabular}
\end{center}
\label{tab_E}
\end{table}

\par We now denote by ${\cal P}_n(t)$ the probability of finding the system in state $n$ at time $t$. Its evolution
 is governed by the master equation
\begin{align}
\frac{\rm d}{{\rm d}t}{\cal P}_n(t)=\sum_{n'(\neq n)}\Big[W_{nn'}{\cal P}_{n'}(t)-W_{n'n}{\cal P}_n(t)\Big] \text{.} 
\label{eq_master_equation}
\end{align}
Transiton rates are given by~\footnote{In their paper [22], Jarzynski and Mazonka only gave the ratio of the transitoin
 rates and adopt the Metropolis algorithm to simulate the dynamics.}
\begin{align}
W_{nn'}=\frac{\beta\Delta U_{n\leftarrow n'}}{\exp(\beta\Delta U_{n\leftarrow n'})-1} \label{eq_rates}
\end{align}
for allowed transitions and $W_{nn'}=0$ for forbidden ones. Here $\Delta U_{n\leftarrow n'}$ is the potential energy change 
associated with the transition from state $n'$ to $n$. For transition $n'\to n$ within one period,the energy difference is
 $\Delta U_{n\leftarrow n'}=U_{n}-U_{n'}$ with energies $U_n$ listed in Table~\ref{tab_E}. However, for those transitions across 
 the boundaries, the potential energy changes are given by
\begin{align}
& \Delta U_{3\leftarrow 1}=fd \text{,} \\
& \Delta U_{1\leftarrow3}=-fd \text{,} \\
& \Delta U_{6\leftarrow 4}=2\alpha+fd \text{,} \\
& \Delta U_{4\leftarrow 6}=-2\alpha-fd \text{.}
\end{align}
$\beta\equiv 1/(k_{\rm B}T)$ is the inverse temperature and $k_{\rm B}$ is Boltzmann's constant. Depending on the the type of 
transition, $T$ may take different values: $T=T_A$ for switches between two modes and $T=T_B$ for jumps between neighbouring sites. 
The rate function~(\ref{eq_rates}) satisfies the detailed balance condition
\begin{align}
W_{nn'}=W_{n'n}\exp(-\beta\Delta U_{n\leftarrow n'}) 
\end{align}
originating from the microscopic reversibility.
 
The master equation~(\ref{eq_master_equation}) with explicit transition rates can be simulated with the Gillespie's algorithm
~\cite{Gillespie_JComputPhys_1976}, which is an explicit method to generate random trajectories for Markov stochastic processes. 
If we define the diagonal elements to be
\begin{align}
W_{nn}\equiv-\sum_{n'(\neq n)}W_{n'n} \text{,}
\end{align}
the master equation~(\ref{eq_master_equation}) can be rewritten into a matrix form as
\begin{align}
\frac{\rm d}{{\rm d}t}{\boldsymbol{\cal P}}={\boldsymbol{\sf L}}\cdot{\boldsymbol{\cal P}} \text{,}
\end{align}
where ${\boldsymbol{\cal P}}=({\cal P}_1,\cdots,{\cal P}_6)^{\rm T}$ and ${\boldsymbol{\sf L}}=[W_{nn'}]$. Each column 
of ${\boldsymbol{\sf L}}$ sums to zero so that it guarantees the conservation of probability.

\section{The Currents and Their Corresponding Affinities}

\par
\begin{figure}
    \begin{minipage}[t]{0.6\hsize}
    \resizebox{1.0\hsize}{!}{\includegraphics{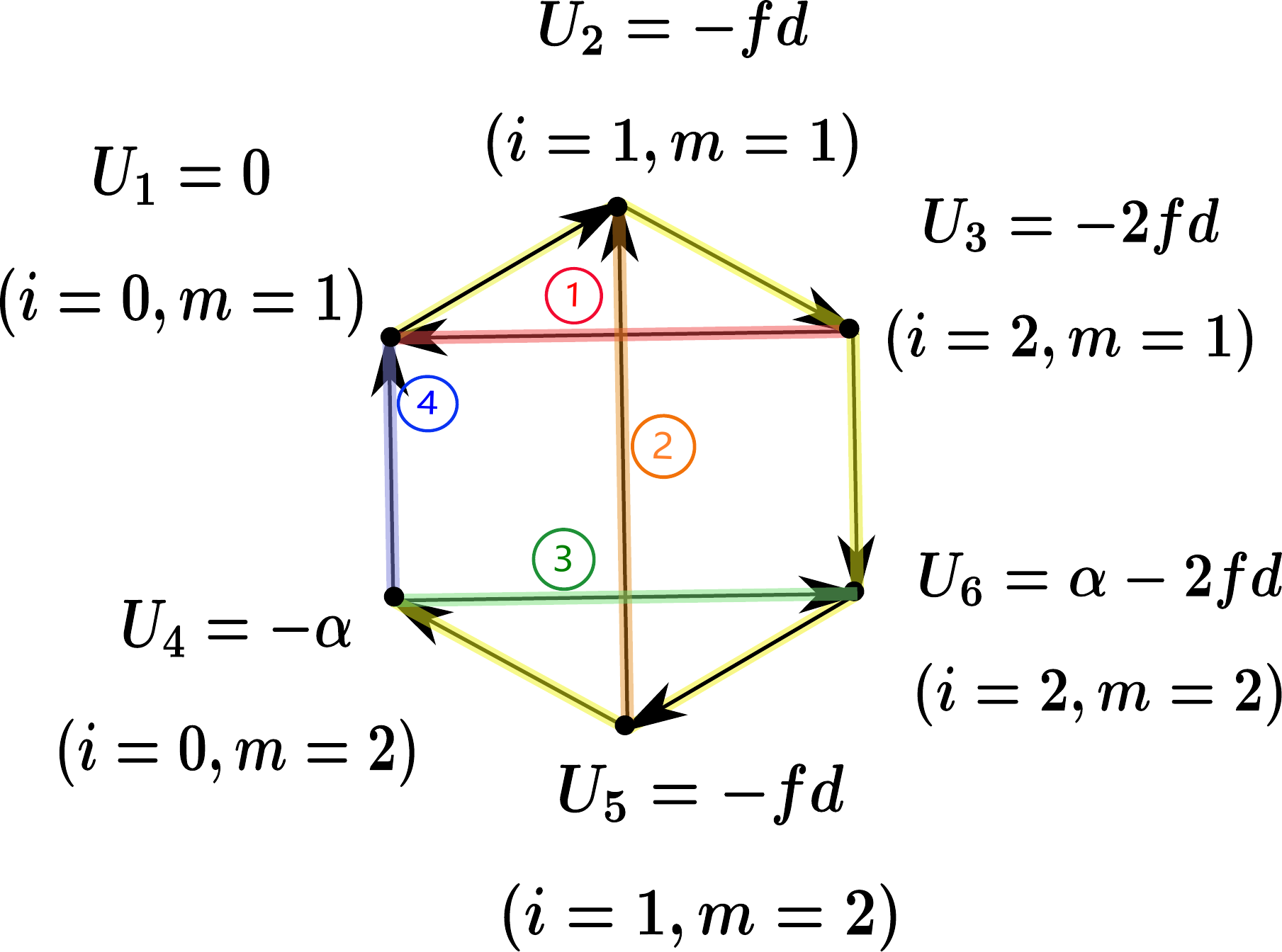}}
    \end{minipage}
    \caption{The graph of the discrete model with vertices representing the distinct states, edges denoting the allowed transitions.
     The subgraph consisting of the thick yellow lines is chosen as the maximal tree. The corresponding chords are numbered with
      \textcircled{1}, \textcircled{2}, \textcircled{3}, \textcircled{4}.}
    \label{fig_graph}
    \end{figure}

 Intuitively we can infer that there are two coupled currents in the ratchet and pawl system, namely the energy flow between 
two heat baths and the displacement of the load. When the system switches the mode, reservoir $A$ exchanges an energy $\alpha$ with the system. Therefore the instantaneous energy current from reservoir A to the system is defined as
\begin{align} 
j_E(t)\equiv\alpha\sum_{k=1}^{+\infty}\Big[ & \delta\left(t-t_k^{1\leftarrow 4}\right)-\delta\left(t-t_k^{4\leftarrow 1}\right) \nonumber \\
& + \delta\left(t-t_k^{6\leftarrow 3}\right)-\delta\left(t-t_k^{3 \leftarrow 6}\right) \Big] \text{,} \label{eq_jE}
\end{align}
where $t_k^{n'\leftarrow n}$ is the $k$-th random transition from state $n$ to $n'$. Similarly, the instantaneous current of 
displacement (velocity) of the load is defined by considering transitions between neighboring sites in both modes, yielding
\begin{widetext}
\begin{align}
j_X(t)\equiv d\sum_{k=1}^{+\infty}\Big[ & \delta\left(t-t_k^{2\leftarrow 1}\right)-\delta\left(t-t_k^{1\leftarrow 2}\right)
 + \delta\left(t-t_k^{3\leftarrow 2}\right)-\delta\left(t-t_k^{2\leftarrow 3}\right) + \delta\left(t-t_k^{1\leftarrow 3}\right)
 -\delta\left(t-t_k^{3\leftarrow 1}\right) \nonumber \\
& +\delta\left(t-t_k^{5\leftarrow 4}\right)-\delta\left(t-t_k^{4\leftarrow 5}\right) + \delta\left(t-t_k^{6\leftarrow 5}\right)
-\delta\left(t-t_k^{5\leftarrow 6}\right) + \delta\left(t-t_k^{4\leftarrow 6}\right)-\delta\left(t-t_k^{6\leftarrow 4}\right) \Big] \text{.} \label{eq_jW}
\end{align}
\end{widetext}
Correspondingly, there are two driving forces or affinities, which can be determined by analyzing the master equation~(\ref{eq_master_equation}) with 
Schnakenberg's network theory~\cite{Schnakenberg_RevModPhys_1976,Faggionato_JStatPhys_2011}. In this theory, a graph is
 associated with the Markov jump process. Vertices represent the states while the edges stand for the allowed 
 transitions between states. The graph for the ratchet system is depicted in Fig.~\ref{fig_graph}. From the so-constructed graph,
  the affinities can be calculated from the transition rates along cyclic paths and their reversals. For example, the cyclic path
   associated with energy $\alpha$ transferred from reservoir A to B and its reversed path could be
\begin{align}
& {\cal C}_E: 1\to 2\to 5\to 4\to 1 \text{,} \\
& {\cal C}_{E}^{r}: 1\to 4\to 5\to 2\to 1 \text{,}
\end{align}
and the affinity is given by
\begin{align}
A_E=\frac{1}{\alpha}\ln\frac{W_{14}W_{45}W_{52}W_{21}}{W_{12}W_{25}W_{54}W_{41}}=\beta_B-\beta_A \text{.}
\end{align}
The prefactor is aimed to set the affinity in the unit of energy. Similarly, for a load displacement of $3d$, the cyclic path
 and its reversal could be
\begin{align}
& {\cal C}_X: 1\to 2\to 3\to 1 \text{,} \\
& {\cal C}_{X}^{r}: 1\to 3\to 2\to 1 \text{.}
\end{align}
The corresponding affinity is
\begin{align}
A_X=\frac{1}{3d}\ln\frac{W_{13}W_{32}W_{21}}{W_{12}W_{23}W_{31}}=f\beta_B \text{.}
\end{align}

We notice that although the transition rates usually depend on the states, the so-obtained affinities only depend
 on the thermodynamic forces which are of physical importance. Since the graph only contains six vertices, it can
  be easily decomposed into four independent cycles which form the fundamental set. Detailed analysis is presented in 
  Appendix~\ref{app_cycles}.

\section{Fluctuation Theorem for the Currents}

\subsection{Full Counting Statistics}

\begin{figure}
  \begin{minipage}[t]{0.8\hsize}
  \resizebox{1.0\hsize}{!}{\includegraphics{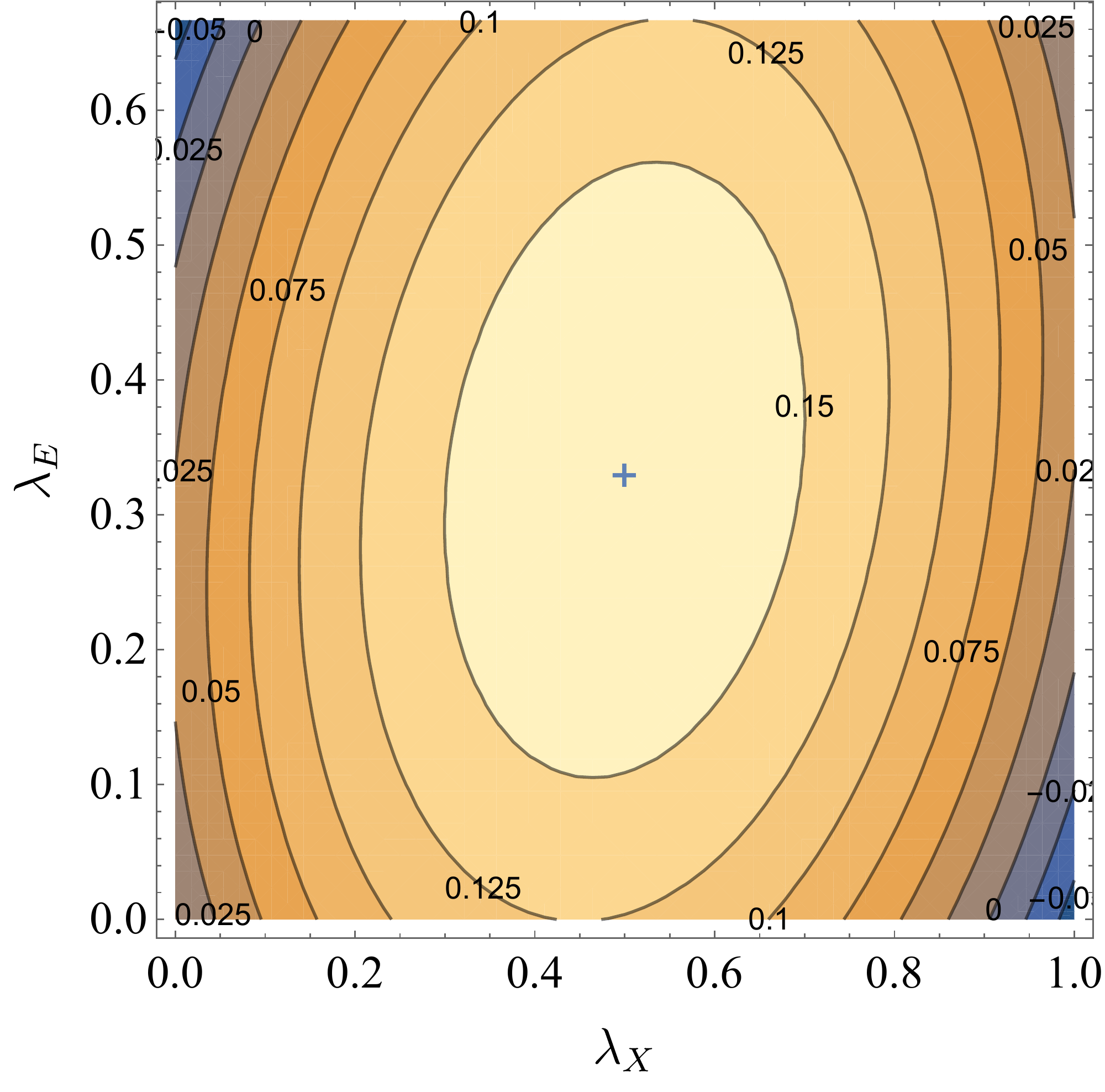}}
  \end{minipage}
  \caption{The contour map of the cumulant generating function $Q(\lambda_X,\lambda_E)$ centering at $(A_X/2=1/2, A_E/2=1/3)$ 
  with $T_A=3, T_B=1, f=1$.}
  \label{fig_Q}
  \end{figure}

\par Giving access to all cumulants of the current, full counting statistics is a powerful theoretical method to study fluctuations.
In order to investigate the full counting statistics, we first define the accumulated energy flowing across the ratchet system and the
accumulated displacement of the load over the time interval $[0,\;t]$
\begin{align}
E(t)\equiv\int_0^tj_E(t')\,{\rm d}t' \text{,}\hspace{0.5cm} X(t)\equiv\int_0^tj_X(t')\,{\rm d}t' \text{.}
\end{align}

\begin{figure*}
\begin{minipage}[t]{0.45\hsize}
\resizebox{1.0\hsize}{!}{\includegraphics{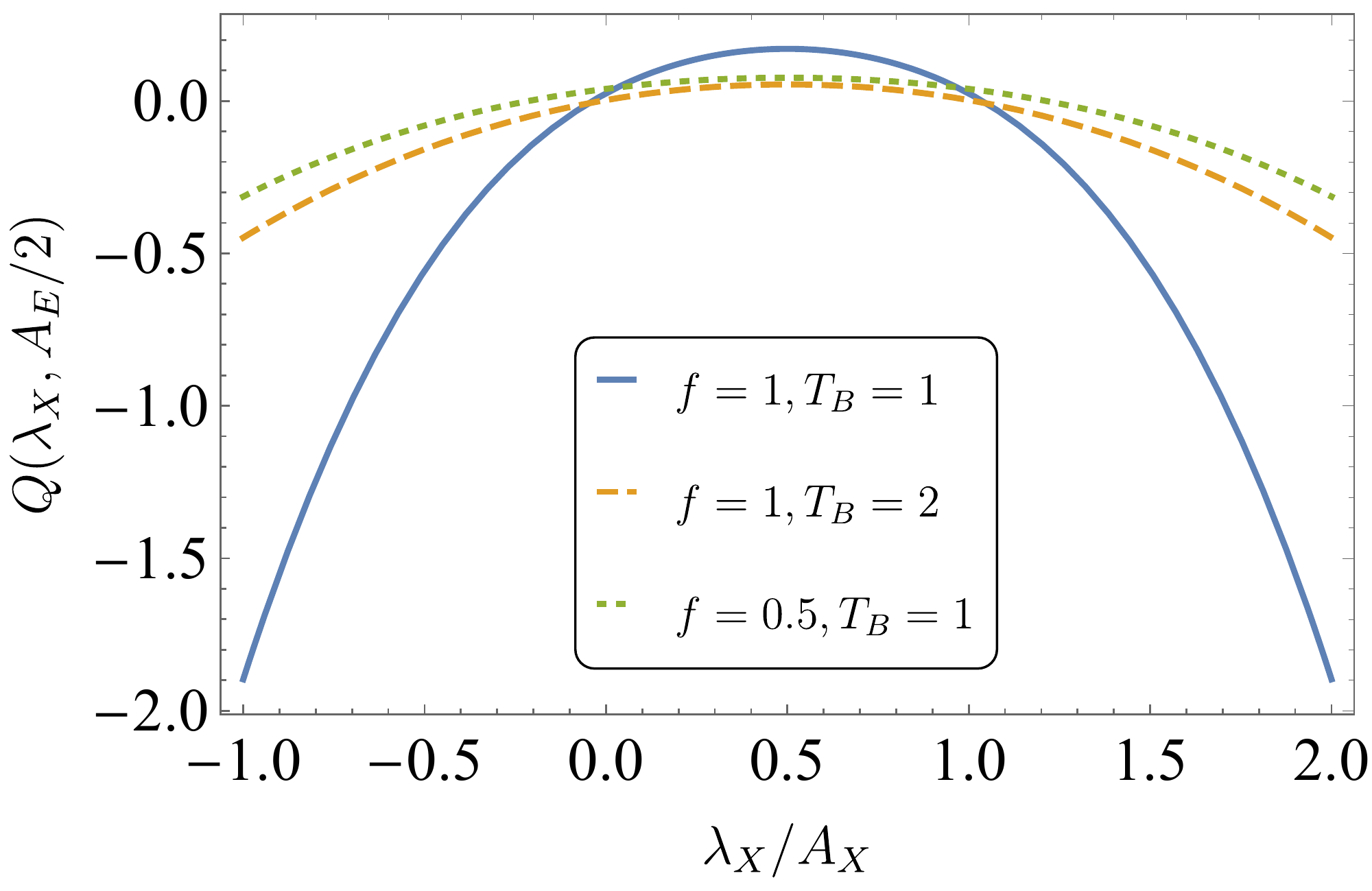}}
\end{minipage}
\begin{minipage}[t]{0.45\hsize}
\resizebox{1.0\hsize}{!}{\includegraphics{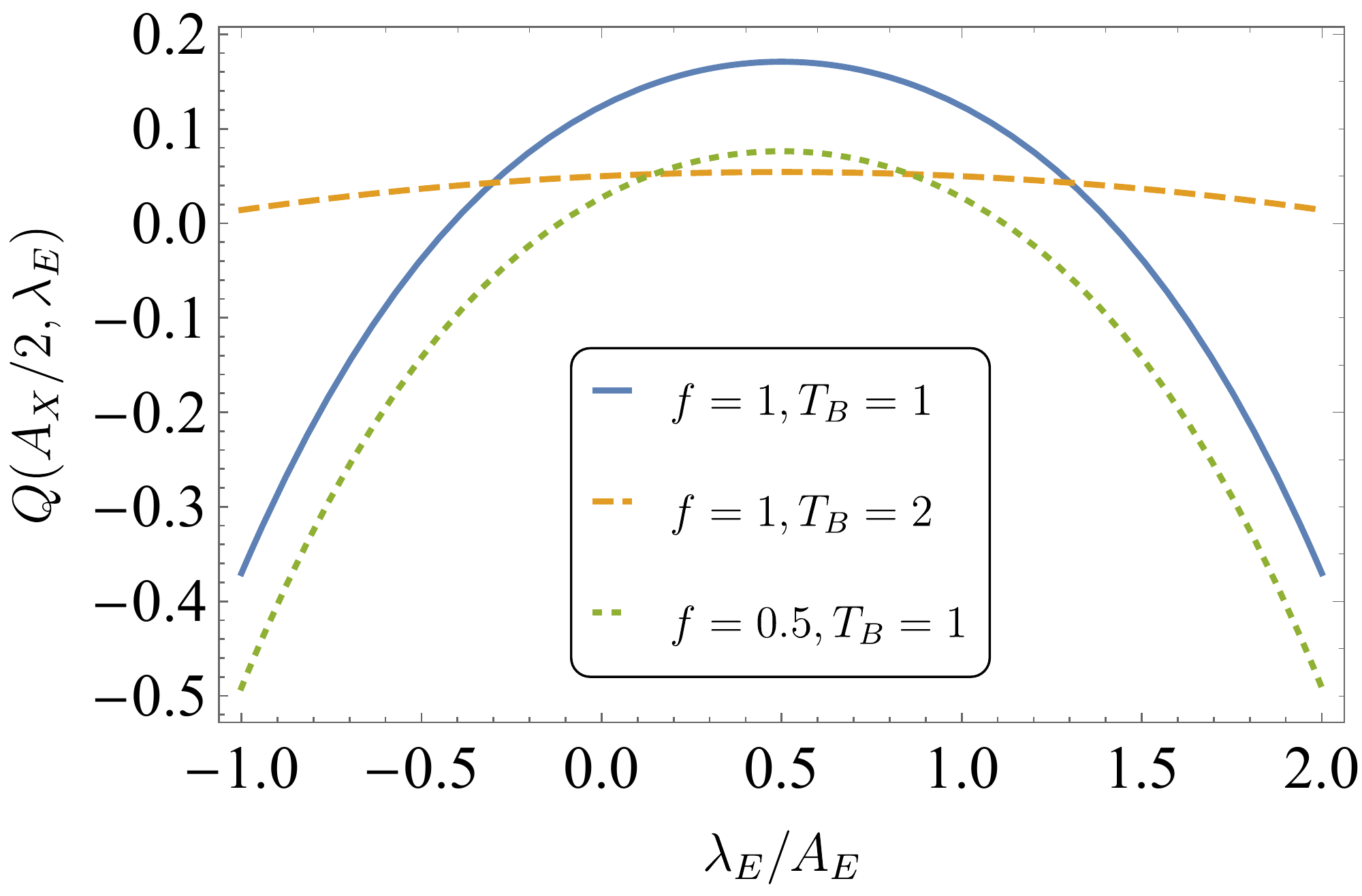}}
\end{minipage}
\caption{The cross sections of the cumulant generating function $Q(\lambda_X,\lambda_E)$. In the left panel, the counting
 parameter is
 $\lambda_E=A_{E}/2$ while in the right panel $\lambda_X=A_{X}/2$. In both panels, $T_A=3$.}
\label{fig_Q12}
\end{figure*}

Then it is eay to obtain the extended master equation for the joint probability ${\cal P}_n(E,X,t)$ of finding the system
 in state $n$ at time $t$ while having absorbed energy $E$ from reservoir A and moved by a distance $X$ (see Appendix~\ref{app_L_lambda} ). By summing over all states,
  we can also define the marginal probability ${\cal P}(E,X,t)\equiv\sum_n{\cal P}_n(E,X,t)$. Note that the energy and  
  displacement take discrete values and change according to $E\to E\pm\alpha$ and $X\to X\pm d$. We further define the
   cumulant generating function as
\begin{align}
Q(\lambda_E,\lambda_X)\equiv\lim_{t\to\infty}-\frac{1}{t}\ln\sum_{E,X}{\cal P}(E,X,t)\,{\rm e}^{-\lambda_EE-\lambda_XX} \text{,}
\end{align}
where $\lambda_E$ and $\lambda_X$ are the counting parameters. The cumulant generating function (CGF) $Q(\lambda_E,\lambda_X)$ 
can be obtained by solving for the leading eigenvalue of operator ${\boldsymbol{\sf L}}_{\lambda_E,\lambda_X}$
\begin{align}
{\boldsymbol{\sf L}}_{\lambda_E,\lambda_X}\cdot{\boldsymbol\Psi}_{\lambda_E,\lambda_X}=-Q(\lambda_E,\lambda_X){\boldsymbol \Psi}_{\lambda_E,\lambda_X} \text{,}
\end{align}
with ${\boldsymbol{\sf L}}_{\lambda_E,\lambda_X}$ being 
\begin{widetext}
\begin{align}
{\boldsymbol{\sf L}}_{\lambda_E,\lambda_X}\equiv\begin{pmatrix}
W_{11} & W_{12}\,{\rm e}^{d\lambda_X} & W_{13}\,{\rm e}^{-d\lambda_X} & W_{14}\,{\rm e}^{-\alpha\lambda_E} & W_{15} & W_{16} \bigstrut \\
W_{21}\,{\rm e}^{-d\lambda_X} & W_{22} & W_{23}\,{\rm e}^{d\lambda_X} & W_{24} & W_{25} & W_{26} \bigstrut \\
W_{31}\,{\rm e}^{d\lambda_X} & W_{32}\,{\rm e}^{-d\lambda_X} & W_{33} & W_{34} & W_{35} & W_{36}\,{\rm e}^{\alpha\lambda_E} \bigstrut \\
W_{41}\,{\rm e}^{\alpha\lambda_E} & W_{42} & W_{43} & W_{44} & W_{45}\,{\rm e}^{d\lambda_X} & W_{46}\,{\rm e}^{-d\lambda_X} \bigstrut \\
W_{51} & W_{52} & W_{53} & W_{54}\,{\rm e}^{-d\lambda_X} & W_{55} & W_{56}\,{\rm e}^{d\lambda_X} \bigstrut \\
W_{61} & W_{62} & W_{63}\,{\rm e}^{-\alpha\lambda_E} & W_{64}\,{\rm e}^{d\lambda_X} & W_{65}\,{\rm e}^{-d\lambda_X} & W_{66} \bigstrut
\end{pmatrix} \label{eq_L_lambda} \text{.}
\end{align}
\end{widetext}
Since the matrix exponential $\exp({\boldsymbol{\sf L}}_{\lambda_E,\lambda_X}t)>0$, the Perron-Frobenius theorem applies and 
the leading eigenvalue $-Q(\lambda_E,\lambda_X)$ of ${\boldsymbol{\sf L}}_{\lambda_E,\lambda_X}$ corresponds to the real 
maximum eigenvalue $\exp[-Q(\lambda_E,\lambda_X)]$ of $\exp({\boldsymbol{\sf L}}_{\lambda_E,\lambda_X}t)$ in magnitude ($t>0$).
 So the ${\boldsymbol\Psi}_{\lambda_E,\lambda_X}$ can be asymptotically evaluated as
\begin{align}
{\boldsymbol\Psi}_{\lambda_E,\lambda_X}\sim_{t\to\infty} \sf exp\left({\boldsymbol{\sf L}}_{\lambda_E,\lambda_X}t\right)\cdot
{\boldsymbol\psi} \text{,}
\end{align}
which is then normalized
\begin{align}
{\boldsymbol\Psi}_{\lambda_E,\lambda_X}\leftarrow\frac{{\boldsymbol\Psi}_{\lambda_E,\lambda_X}}{\sqrt{{\boldsymbol\Psi}_{\lambda_E,\lambda_X}^{\rm T}\cdot{\boldsymbol\Psi}_{\lambda_E,\lambda_X}}} \text{.}
\end{align}
Here ${\boldsymbol\Psi}_{\lambda_E,\lambda_X}$ is the eigenvector corresponding to the leading eigenvalue of ${\boldsymbol{\sf L}}_{\lambda_E,\lambda_X}$, and
$\boldsymbol{\psi}$ is the randomly chosen distribution supposed to include the desired component. 
 The matrix exponential can be computed using Pad\'e approximation~\footnote{In mathematics, Pad\'e approximants form 
a particular type of rational fraction approximation to the value of the function. They often give a better approximation 
of the function than truncating its Taylor series and may still work where the Taylor series does not converge. For 
these reasons Pad\'e approximants are extensively used in computer calculations.}~\cite{Higham_book_2008}. The cumulant generating function is calculated as

\begin{align}
Q(\lambda_E,\lambda_X)=-{\boldsymbol\Psi}_{\lambda_E,\lambda_X}^{\rm T}\cdot{\boldsymbol{\sf L}}_{\lambda_E,\lambda_X}\cdot{\boldsymbol\Psi}_{\lambda_E,\lambda_X} \text{.}
\end{align}
The stationary distribution is given by ${\boldsymbol{\cal P}}_{\rm st}={\boldsymbol\Psi}_{0,0}$. Fig.~\ref{fig_Q} shows in
 contour map the cumulant generating function in the plane of the counting parameters $\lambda_E$ and $\lambda_X$. Two cross 
 sections are plotted in Fig.~\ref{fig_Q12}.

\subsection{Fluctuation Theorem}

\par We find that matrix ${\boldsymbol{\sf L}}_{\lambda_E,\lambda_X}$~(\ref{eq_L_lambda}) exhibits the symmetry relation
\begin{align}
{\boldsymbol{\sf M}}^{-1}\cdot{\boldsymbol{\sf L}}_{\lambda_E,\lambda_X}\cdot{\boldsymbol{\sf M}}={\boldsymbol{\sf L}}_
{A_E-\lambda_E,A_X-\lambda_X}^{\rm T}  \label{eq_L_symmetry}
\end{align}
under similarity transformation with matrix ${\boldsymbol{\sf M}}$~\cite{Lau_PhysRevLett_2007,Lacoste_PhysRevE_2009,Lacoste_PhysRevE_2008}
\begin{align}
{\boldsymbol{\sf M}}=\begin{pmatrix}
1 & 0 & 0 & 0 & 0 & 0 \bigstrut \\
0 & 1 & 0 & 0 & 0 & 0 \bigstrut \\
0 & 0 & 1 & 0 & 0 & 0 \bigstrut \\
0 & 0 & 0 & {\rm e}^{\alpha/{dT_B}} & 0 & 0 \bigstrut \\
0 & 0 & 0 & 0 & 1 & 0 \bigstrut \\
0 & 0 & 0 & 0 & 0 & {\rm e}^{-\alpha/{dT_B}} \bigstrut \\
\end{pmatrix} \text{.}
\end{align}
As similarity transformation of the matrix does not change the eigenvalues, the relation~(\ref{eq_L_symmetry}) implies 
\begin{align}
Q({\lambda_E,\lambda_X})=Q(A_E-\lambda_E,A_X-\lambda_X) \text{,} \label{eq_GC}
\end{align}
which is known as Gallavotti-Cohen symmetry~\cite{Gallavotti_PhysRevLett_1995,
Gallavotti_PhysRevLett_1995,Faggionato_JStatPhys_2011}. This symmetry for the CGF can be seen in 
Fig.~\ref{fig_Q} and Fig.~\ref{fig_Q12}. The CGF is directly related to the large deviation 
function through Legendre-Fenchel transform~\cite{Touchette_PhysRep_2009}
\begin{align}
I(e,x)=\max_{\lambda_E,\lambda_X}\Big[Q(\lambda_E,\lambda_X)-\left(\lambda_Ee+\lambda_Xx\right)\Big] \text{.}
\end{align}
$I(e,x)$ is the large deviation rate function of the probability ${\cal P}(E=et,X=xt,t)$ defined as follows
\begin{align}
{\cal P}(E=et,X=xt,t)\sim_{t\to\infty} {\rm e}^{-tI(e,x)} \text{.}
\end{align}
The Gallavotti-Cohen symmetry relation~(\ref{eq_GC}) immediately implies
\begin{align}
I(-e,-x)-I(e,x)=A_Ee+A_Xx \text{,}
\end{align}
leading all the way to
\begin{align}
  \frac{{\cal P}(E, X, t)}{{\cal P}(-E, -X, t)}\simeq_{t\to\infty} {\rm e}^{A_EE+A_XX} \text{,} \label{eq_FT}
\end{align}
which is the usual form of fluctuation theorem.
This relation can also be derived based on the cycle theory (See Appendix.~\ref{app_cycles}).
\par What's more, the similarity between  ${\boldsymbol{\sf L}}_{\lambda_E,\lambda_X}$ and ${\boldsymbol{\sf L}}_
{A_E-\lambda_E,A_X-\lambda_X}^{\rm T}$ implies that all eigenvalues are the same, leading to a finite-time 
fluctuation theorem
\begin{align}
\left\langle{\rm e}^{-\lambda_EE-\lambda_XX}\right\rangle=\left\langle{\rm e}^{-(A_E-\lambda_E)E-(A_X-\lambda_X)X}\right\rangle\text{.}
\label{eq_finite_time_FT}
\end{align}
This coincides with the fluctuation theorem of entropy production~\cite{Seifert_PhysRevLett_2005}.

\subsection{Thermodynamic Entropy Production}

\par For the ratchet system in nonequilibrium steady state, the mean currents of energy and displacement can be evaluated as
\begin{align}
& J_E=\lim_{t\to\infty}\frac{1}{t}\sum_{E,X}{\cal P}(E,X,t)E \text{,} \\
& J_X=\lim_{t\to\infty}\frac{1}{t}\sum_{E,X}{\cal P}(E,X,t)X \text{.}
\end{align}
The thermodynamic entropy production rate~\cite{Prigogine_1967, Callen_1998} in units of Boltzmann's constant is identified 
as the sum of products of the affinities and the corresponding mean currents. According to the fluctuation theorem~(\ref{eq_FT}), it
can be expressed as
\begin{align}
\frac{1}{k_{\rm B}}\frac{{\rm d}_{\rm i}S}{{\rm d}t} & =A_EJ_E+A_XJ_X \nonumber \\
& =\lim_{t\to\infty}\frac{1}{t}\sum_{E,X}{\cal P}(E, X, t)\ln\frac{{\cal P}(E, X, t)}{{\cal P}(-E, -X, t)} \text{,} 
\label{eq_entropy_production}
\end{align}
in terms of the Kullback-Leibler divergence of ${\cal P}(E,X,t)$ and ${\cal P}(-E,-X,t)$. The term on the
 r.h.s. of Eq.~(\ref{eq_entropy_production}) is always non-negative, which is in accordance with the second law of 
 thermodynamics. Also, Eqs.~(\ref{eq_FT},\ref{eq_finite_time_FT}) indicate that there is a nonzero probability of observing the "violation"
of the second law in the Feynman's ratchet, i.e.,$A_EE+A_XX<0$. But in the Feynman's ratchet, the probability of observing such a "violation" becomes negligibly small 
in the long time limit since both $E$ and $X$ are proportional to $t$. These results obviously agree
with our intuition about Feynman's ratchet.

\section{Symmetry Relations for the Response Coefficients}

\par It has been shown in Refs.~\cite{Andrieux_NewJPhys_2009,Barbier_JPhysA_2018,Barbier_PhysRevE_2020,
Barbier_JPhysA_2020,Gu_JStatMech_2020,Gu_PhysRevE_2019,Derrida_JStatMech_2007,CholeYaGao_JChemPhys_2019} that the Gallavotti-Cohen 
symmetry relation provides a unified framework from
 which one can derive relations between response coefficients to an arbitrary order (See Eqs.~(\ref{eq_FD}-\ref{eq_MRR}) below).
 In previous sections we fix the affinities $A_X, A_E$. In this section, we will evaluate the response properties for different values of afffinites.
 Hence, the CGF $Q(\lambda_X,\lambda_E)$ should be rewrittened as
 $Q(\boldsymbol{\lambda},\bf A)$. Here $\boldsymbol{\lambda}=\set{\lambda_1,\lambda_2},\bf A$$= \left\{A_1, A_2\right\}$, $1=X, 2=E.$
  The cumulants can be evaluated by taking successive derivatives of the CGF with respect to the
  counting parameters. The average currents and diffusivities are
   given by
\begin{align}
J_i({\bf A})=\left.\frac{\partial Q({\boldsymbol\lambda};{\bf A})}{\partial\lambda_i}\right\vert_{{\boldsymbol\lambda}={\bf 0}} \text{,}
\end{align} 
\begin{align}
D_{ij}({\bf A})=\left.-\frac{1}{2}\frac{\partial^2 Q({\boldsymbol\lambda};{\bf A})}{\partial\lambda_i\partial\lambda_j}\right\vert_{{\boldsymbol\lambda}={\bf 0}} \text{.}
\end{align}
In the mean time, we can expand the mean currents in the power series of afffinites as 
\begin{align}
J_i({\bf A})=\sum_jL_{i,j}A_j+\frac{1}{2}\sum_{j,k}M_{i,jk}A_jA_k+\cdots 
\end{align}
in terms of the linear and nonlinear response coefficients defined by~\cite{Andrieux_NewJPhys_2009}
\begin{align}
& L_{i,j}\equiv\left.\frac{\partial^2 Q({\boldsymbol\lambda};{\bf A})}{\partial\lambda_{i}\partial A_{j}}\right\vert_{{\boldsymbol\lambda}={\bf A}={\bf 0}} \text{,}    \\
& M_{i,jk}\equiv\left.\frac{\partial^{3} Q({\boldsymbol\lambda};{\bf A})}{\partial\lambda_{i}\partial A_{j}\partial A_{k}}\right\vert_{{\boldsymbol\lambda}={\bf A}={\bf 0}} \text{,}
\end{align}
where we notice $M_{i,jk}=M_{i,kj}$ by definition. From the Gallavotti-Cohen symmetry relation~(\ref{eq_GC}), we can easily
 deduce the fluctuation-dissipation theorem~\cite{Kubo_RepProgPhys_1966}

\begin{align}
L_{i,j}=D_{ij}({\bf A=\bf 0}) \text{,} \label{eq_FD}
\end{align}

and Onsager reciprocal relations~\cite{Onsager_PhysRev_1931b, Casimir_RevModPhys_1945,Andrieux_NewJPhys_2009,Jarzynski_PhysRevE_1999}

\begin{align}
L_{i,j}=L_{j,i} \text{.} \label{eq_Onsager}
\end{align}

We can also prove the relations characterizing the second order response properties~\cite{Andrieux_NewJPhys_2009}
\begin{align} 
M_{i,jk}=R_{ij,k}+R_{ik,j} \text{,} \label{eq_MRR}
\end{align}
where $R_{ij,k}$ is defined as the linear response coefficient of the diffusivity around equilibrium,
\begin{align}
R_{ij,k}\equiv\left.\frac{\partial D_{ij}}{\partial A_k}\right\vert_{{\bf A}={\bf 0}}=-\frac{1}{2}\left.\frac{\partial
^3Q({\boldsymbol\lambda};{\bf A})}{\partial\lambda_i\partial\lambda_j\partial A_k}\right\vert_{{\boldsymbol\lambda}
={\bf A}={\bf 0}} \text{.} \label{eq_R}
\end{align}
We notice that Onsager reciprocal relation~(\ref{eq_Onsager}) is the direct consequence of the fluctuation-dissipation
 relations~(\ref{eq_FD}) together with the symmetry of the diffusivities $D_{ij}=D_{ji}$. Higher order relations can 
 be obtained likewise.

\par
\begin{figure}
\begin{minipage}[t]{0.8\hsize}
\resizebox{1.0\hsize}{!}{\includegraphics{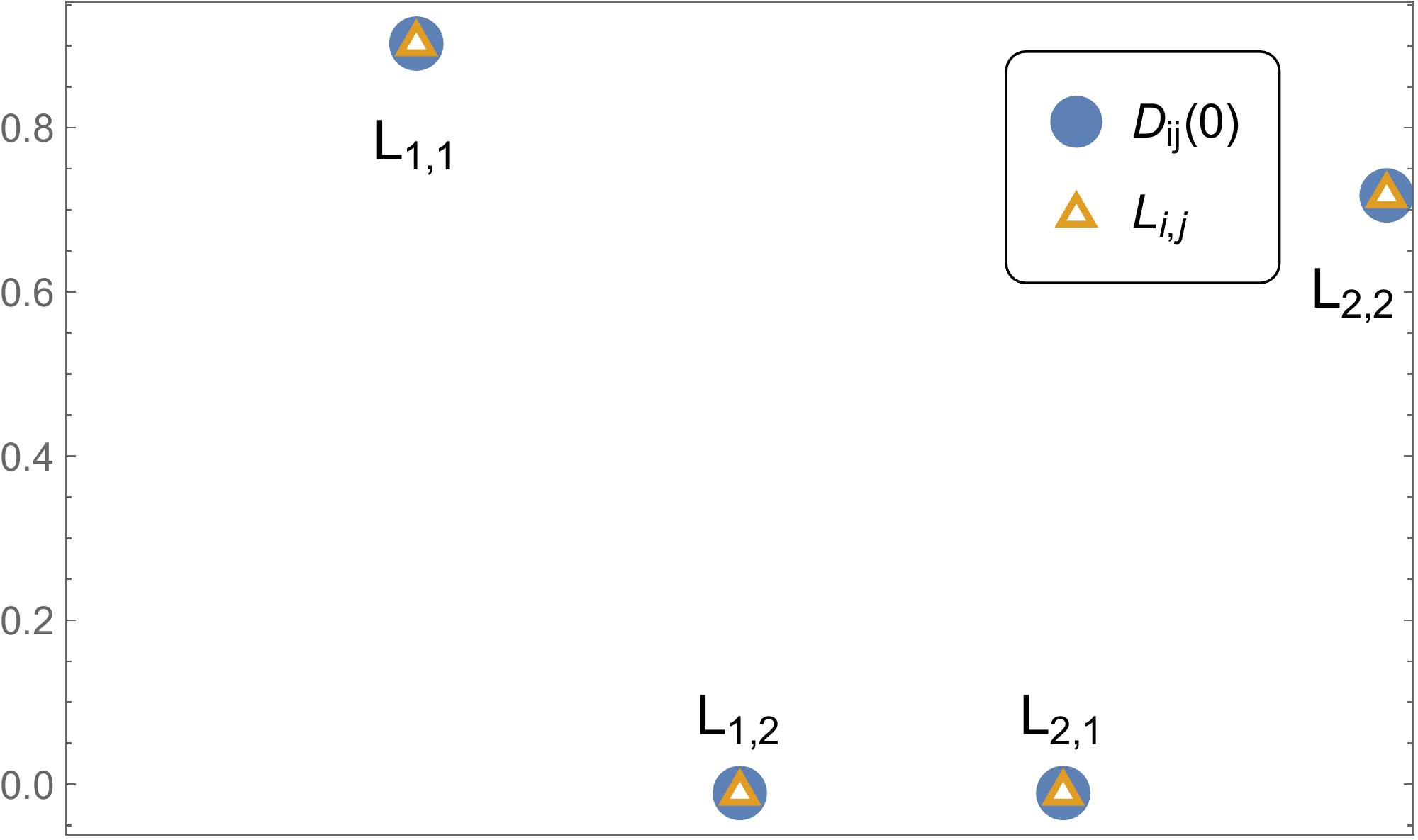}}
\end{minipage}
\caption{The comparison between linear response coefficients $L_{i, j}$ (orange triangles) and diffusivities 
in equilibrium $D_{ij}({\bf 0})$ (blue circles). The parameters are taken to be the same as those
 in Fig.~\ref{fig_Q}.}
\label{fig_LD}
\end{figure}

\begin{figure}
\begin{minipage}[t]{0.8\hsize}
\resizebox{1.0\hsize}{!}{\includegraphics{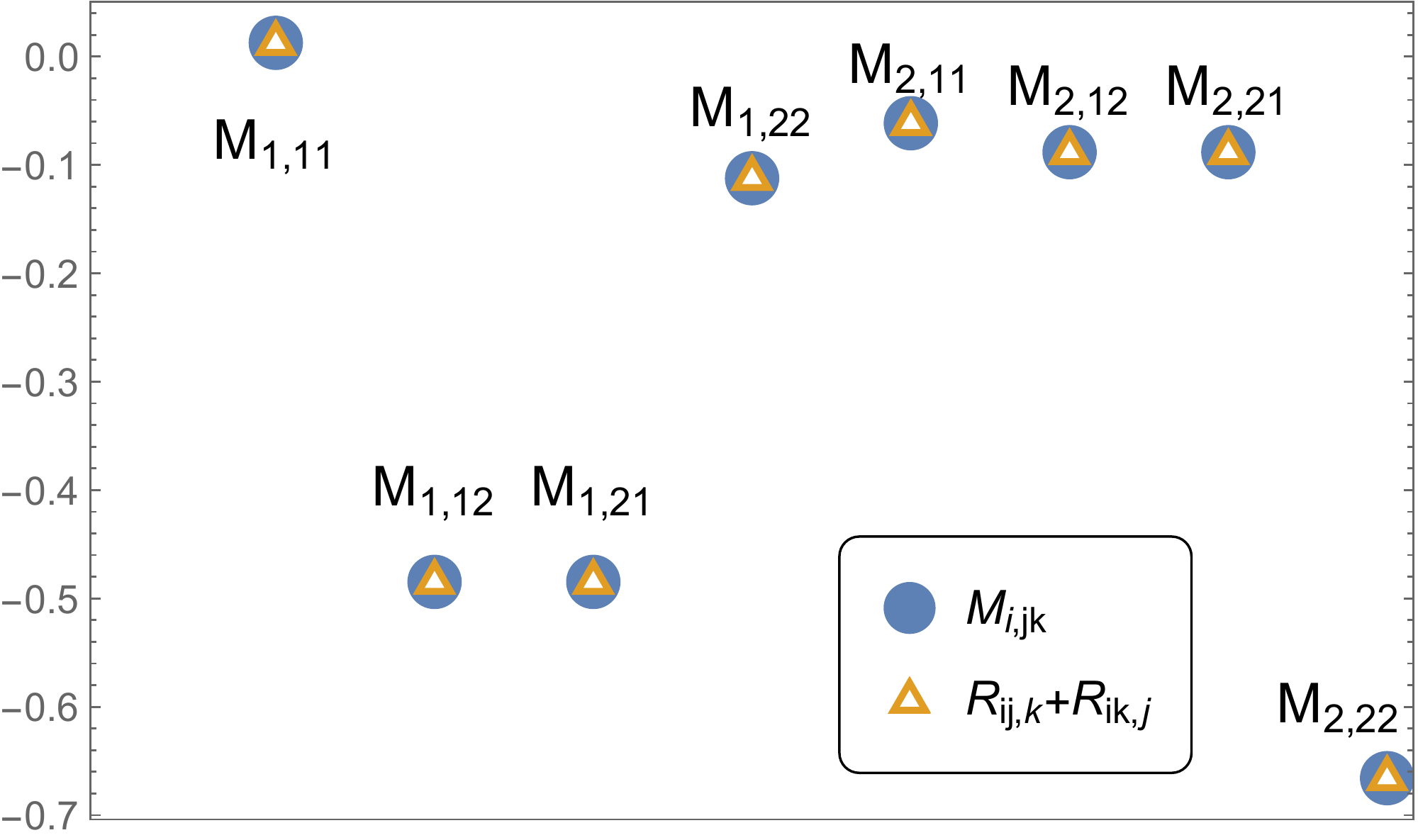}}
\end{minipage}
\caption{The comparison between the second-order response coefficients $M_{i,jk}$ and $R_{i,jk}+R_{i,kj}$, where
 $R_{i,jk}$ is defined by Eq.~(\ref{eq_R}). Same parameters as in Fig.~\ref{fig_Q}.}
\label{fig_MR}
\end{figure}

 We do a numerical test of the relations~(\ref{eq_FD})-(\ref{eq_MRR}) in the ratchet system. For this purpose, 
we first calculate the CGF $Q(\lambda_X,\lambda_E;A_X,A_E)$ for several points around
 $\lambda_X=\lambda_E=A_X=A_E=0$. Then we perform the Lagrange interpolation to obtain the multivariate
  polynomial approximating the CGF. Finally, the values of $L_{i,j}$, $D_{ij}$, $M_{i,jk}$ and
   $R_{ij,k}$ for $i,j,k=1,2$ $(1$ for $X, 2$ for $E )$ are calculated by taking derivatives according to their definitions.
    The agreement between the related quantities is clearly demonstrated in Figs.~\ref{fig_LD} and~\ref{fig_MR}. Thus for 
    the ratchet system, the validity of predictions of fluctuation theorems is tested in both linear and nonlinear regimes.

\section{Conclusion and Perspectives}

\par In this article, we investigate the full counting statistics of Feynman's ratchet and show that Gallavotti-Cohen fluctuation theorem holds for the currents in the discrete model of Feynman's 
ratchet. The Markovian stochastic jump process is described by the master equation, with transition rates determined by energy
 difference. Detailed balance condition is guaranteed so that the dynamics is compatible with the underlying law of
  microreversibility. Feynman's ratchet and pawl system is driven out of equilibrium by the temperature difference between 
  two heat baths and also by the external load. Correspondingly, there are two physical currents coupled to each other.
   By perfoming full counting  statistics analysis of the currents, we obtain the CGF which manifests a symmetry relation. This symmetry quantitatively
   characterizes the fluctuation and the probablistic "violation" of the second law by the Feynman's ratchet and hence deepens our understanding
   about fluctuating properties of the famous model from the qualitative way to the quantitative way. In addition, we point out
   that the currents exhibit linear and nonlinear response properties in the ratchet system, as predicted by the fluctuation
    theorem. We numerically verify the relation between the linear response coefficients and the equilibrium diffusivities.
     Further we verify in detail that the second-order response coefficients are related to the linear
     response of the diffusivities. Higher order relations can be obtained likewise.

\par As the six-state model of Feynman's ratchet is simple enough, it can be regarded as a basis for studying more sophisticated
 systems, e.g., spatially extended models. Considering that the pawl stochastically switches between two modes, a feedback 
 control can be applied  to determine whether or not the external load is to be attached. In this way, a fluctuation theorem
  involving information flow could  possibly be explored.

\section*{acknowledgments}

\par H. T. Quan acknowledges Christopher Jarzynski for stimulating discussions at the early stage of this work. Financial support from National Science Foundation of China under the Grant No. 12147162, 11775001 and 11825501 
is acknowledged.

\appendix

\section{Cycle Decomposition}\label{app_cycles}

\par Schnakenberg found that~\cite{Schnakenberg_RevModPhys_1976} stochastic processes described by the master equation 
can be investigated by carrying out graph analysis and that the nonequilibrium constraints exerted on a system are related to
 affinities of the cycles in the graph. This mathematical tool is applied here to study fluctutaion theorem for the currents 
 -- energy flow and load displacement.

\par In the basic graph $G$, the vertices represent the distinct states while the edges stand for the allowed transitions 
between states. Then we choose a maximal tree $T(G)$ which is a covering subgraph of $G$. The requirements for a maximal 
tree are listed as follows:
\begin{itemize}
\item $T(G)$ covers all the vertices of $G$ and all the edges of $T(G)$ belong to $G$;
\item $T(G)$ must be connected;
\item $T(G)$ contains no circuit.
\end{itemize}
The edges $l$ of graph $G$ that does not belong to $T(G)$ are refered to as chords of maximal tree $T(G)$. By respectively
 adding one chord to the maximal tree, we get exactly one closed circuit at a time in the resulting subgraph $T(G)+l$. 
 The set of circuits is called a fundamental set $\{C_{l}\}$. As marked in Fig.~\ref{fig_graph}, by choosing the maximal 
 tree $T(G)$ in yellow, the remaining four edges automatically become its chords. It has been identified that there are four
  independent cycles in the fundamental set $\{C_l\}=\{C_{\rm I},\,C_{\rm II},\,C_{\rm III},\,C_{\rm IV}\}$
\begin{align}
& C_{\rm I}=\left(e_{1\to 2},\,e_{2\to 3},\,e_{3\to 1}\right) \text{,} \\
& C_{\rm II}=\left(e_{2\to 3},\,e_{3\to 6},\,e_{6\to 5},\,e_{5\to 2}\right) \text{,} \\
& C_{\rm III}=\left(e_{4\to 6},\,e_{6\to 5},\,e_{5\to 4}\right) \text{,} \\
& C_{\rm IV}=\left(e_{1\to 2},\,e_{2\to 3},\,e_{3\to 6},\,e_{6\to 5},\,e_{5\to 4},\,e_{4\to 1}\right) \text{,}
\end{align}
where $e_{i\to j}$ denotes the directional edge representing the transition from state $i$ to $j$. Schnakenberg's graph
 analysis tells us that each cycle $C$ in the basic graph can be expressed as linear combination of cycles in the fundamental
  set. In the long-time limit, the initial state becomes irrelevant and the whole trajectory forms a closed cycle $C(t)$. So we have
\begin{equation}
C(t)=a_{\rm I}(t)C_{\rm I}+a_{\rm II}(t)C_{\rm II}+a_{\rm III}(t)C_{\rm III}+a_{\rm IV}(t)C_{\rm IV} \text{.}
\end{equation}

\par Now we apply Schnakenberg's network theory to prove that Gallavotti-Cohen symmetry exists in the CGF of the energy flow 
and load displacement. The load displacement can be evaluated by counting the jump times between states in both modes
\begin{align}
X(t)= d \Big[ & S_{1\to 2}+S_{2\to 3}+S_{3\to 1}  \nonumber \\
 &\quad\quad +S_{4\to 5}+S_{5\to 6}+S_{6\to 1}\Big]C(t) \text{,} \label{eq_X}
\end{align}
where the operator $S_{i\to j}$ counts the jump times across the edge $e_{i\to j}$. Such linear operators $S_{i\to j}$ 
separately act on independent cycles
\begin{align}
& S_{1\to 2}C(t)=a_{\rm I}(t)+a_{\rm IV}(t) \text{,} \label{eq_S12}\\
& S_{2\to 3}C(t)=a_{\rm I}(t)+a_{\rm II}(t)+a_{\rm IV}(t) \text{,} \\
& S_{3\to 1}C(t)=a_{\rm I}(t) \text{,} \\
& S_{4\to 5}C(t)=-a_{\rm III}(t)-a_{\rm IV}(t) \text{,} \\
& S_{5\to 6}C(t)=-a_{\rm II}(t)-a_{\rm III}(t)-a_{\rm IV}(t) \text{,} \\
& S_{6\to 4}C(t)=-a_{\rm III}(t)\text{.} \label{eq_S64}
\end{align}
Substituding (\ref{eq_S12}-\ref{eq_S64}) into (\ref{eq_X}) we have
\begin{align}
X(t)=d\Big[3a_{\rm I}(t)-3a_{\rm III}(t)\Big] \text{.}
\end{align}

The energy transferred from the reservoir $A$ to the system can be similarly evaluated by considering
 only the transitions between two modes
\begin{align}
E(t)=\alpha\Big[S_{4\to 1}+S_{3\to 6}\Big]C(t)=\alpha\Big[a_{\rm II}(t)+2a_{\rm IV}(t)\Big] \text{.}
\end{align}
The CGF for the energy flow and load displacement can be written as
\begin{widetext}
\begin{align}
Q(\lambda_E,\lambda_X) & =\underset{t\rightarrow\infty}{\lim}-\frac{1}{t}\ln\left\langle{\rm e}^{-\lambda_EE(t)-
\lambda_XX(t)}\right\rangle \nonumber \\
 & =\lim_{t\to\infty}-\frac{1}{t}\ln\left\langle{\rm e}^{-3\lambda_Xa_{\rm I}(t)-\lambda_Ea_{\rm II}(t)+3\lambda_Xa_
 {\rm III}(t)-2\lambda_Ea_{\rm IV}(t)}\right\rangle \text{.}
\end{align}
Recalling the fluctuation theorem for the currents crossing the chords proved by Andrieux and Gaspard~\cite{Andrieux_JStatPhys_2007}, the CGF of independent currents along the chords 
\begin{align}
q(\{\lambda_l\})=\lim_{t\to\infty}-\frac{1}{t}\ln\left\langle{\rm e}^{\sum_l\lambda_la_l(t)}\right\rangle
\end{align}
exhibits the following symmetry
\begin{align}
q(\{\lambda_l\})=q(\{A_l-\lambda_l\}) \text{,}
\end{align}
where $A_l$ is the affinity of the independent cycle in the fundamental set and the notation $\{\lambda_l\}$ is short for
 $\{\lambda_{\rm I},\cdots,\lambda_{\rm IV}\}$. In this case, the affinities $\{A_l\}$ can be calculated by taking the log of
  ratio of transition rate products in both directions
\begin{equation}
A_{\rm I}=\frac{1}{d}\ln\frac{W_{13}W_{32}W_{21}}{W_{12}W_{23}W_{31}}=3f\beta_B \text{,}
\end{equation}
and similarly 
\begin{align}
& A_{\rm II}=\beta_B-\beta_A \text{,} \\
& A_{\rm III}=-3f\beta_B \text{,} \\
& A_{\rm IV}=2\left(\beta_B-\beta_A\right) \text{.}
\end{align}
When $\{C_l\}$ is a fundamental set, the coefficients $\{a_l(t)\}$ in cycle decomposition equal the integrated current
 along the chords, i.e., the number of times crossing each chord. Then the CGF for joint energy and displacement flow
  $Q(\lambda_E,\lambda_X)$ can be mapped onto $q(\{\lambda_l\})$ as
\begin{align}
Q(\lambda_E,\lambda_X) & =q\left(3\lambda_X,\lambda_E,-3\lambda_X,2\lambda_E\right) \nonumber \\
 & =q\left(A_{\rm I}-3\lambda_X,A_{\rm II}-\lambda_E,A_{\rm III}+3\lambda_X,A_{\rm IV}-2\lambda_E\right) \nonumber \\
 & =q\left(3f\beta_B-3\lambda_X,\beta_B-\beta_A-\lambda_E,-3f\beta_B+3\lambda_X,2\beta_B-2\beta_A-2\lambda_E\right) \nonumber \\
 & =Q(A_E-\lambda_E,A_X-\lambda_X) \text{,}
\end{align}
which gives the Gallavotti-Cohen symmetry~(\ref{eq_GC}).
\end{widetext}

\begin{widetext}
\section{Derivation of Eq.~(\ref{eq_L_lambda})}\label{app_L_lambda}
\par The generating function for the joint energy flow (from reservoir A to the system) and the displacement is defined as
\begin{align}
  \boldsymbol{\Psi}_{\lambda_E,\lambda_X} (t) =\underset{n}{\sum }(\boldsymbol{\Psi}_{\lambda_E,\lambda_X})_{n} (t) \text{,}
\end{align}
\begin{align}
  (\boldsymbol{\Psi}_{\lambda_E,\lambda_X})_{n} (t)&=\int \mathrm{d} E \mathrm{d} X \mathrm{e}^{-\lambda_{E}E-\lambda_{X}X}{\cal P}_{n}(E,X,t),
\end{align}
where ${\cal P}_{n}(E,X,t)$ represents the probability of finding the system in state $n$ at time $t$ while having absorbed 
energy $E(t)$ from reservoir A and moved by a displacement $X(t)$. The displacement of the particle and the heat absorbed from 
reservoir A during a transition from state $m$ to state $n$ is given by $\Delta X_{nm}$ and $\Delta E_{nm}$.
\par In a small time interval $\tau$ the variation of ${\cal P}_{n}(E,X,t)$ is given by

  \begin{align}
    {\cal P}_{n}(E,X,t+\tau)&\approx{\cal P}_{n}(E,X,t)+\tau\underset{m(\neq n)}{\sum}\Bigl[W_{nm}{\cal P}_{m}
    (E-\Delta E_{nm},X-\Delta X_{nm},t)-W_{mn}{\cal P}_{n}(E,X,t)\Bigr]\\
   & ={\cal P}_{n}(E,X,t)+\tau\underset{m(\neq n)}{\sum}\biggl\{W_{nm}\underset{k,l=0}{\sum}\Bigl[\frac{(-\Delta E_{nm})^{k}}
   {k!}\frac{(-\Delta X_{nm})^{l}}{l!}\frac{\partial^{k+l}{\cal P}_{m}(E,X,t)}{\partial E^{k}\partial X^{l}}\Bigr]-W_{mn}
   {\cal P}_{n}(E,X,t)\biggr\}.
  \end{align}
  The joint distribution function ${\cal P}_{n}(E,X,t)$ evolves as
  \begin{align}
    \frac{\partial{\cal P}_{n}(E,X,t)}{\partial t}=\underset{m(\neq n)}{\sum}\biggl\{W_{nm}\underset{k,l}{\sum}
    \Bigl[\frac{(-\Delta E_{nm})^{k}}{k!}\frac{(-\Delta X_{nm})^{l}}{l!}\frac{\partial^{k+l}{\cal P}_
    {m}(E,X,t)}{\partial E^{k}\partial X^{l}}\Bigr]-W_{mn}{\cal P}_{n}(E,X,t)\biggr\}  \label{eq_extended master}
  \end{align}
  with $(\Delta E)_{nm}$ and $(\Delta X)_{nm}$ respectively given by
  \begin{align}
    (\Delta X)_{nm}=d\left(\begin{array}{cccccc}
      0 & -1 & 1 & 0 & 0 & 0\\
      1 & 0 & -1 & 0 & 0 & 0\\
      -1 & 1 & 0 & 0 & 0 & 0\\
      0 & 0 & 0 & 0 & -1 & 1\\
      0 & 0 & 0 & 1 & 0 & -1\\
      0 & 0 & 0 & -1 & 1 & 0
      \end{array}\right)\ \text{,}\ (\Delta E)_{nm}=\alpha\left(\begin{array}{cccccc}
      0 & 0 & 0 & 1 & 0 & 0\\
      0 & 0 & 0 & 0 & 0 & 0\\
      0 & 0 & 0 & 0 & 0 & -1\\
      -1 & 0 & 0 & 0 & 0 & 0\\
      0 & 0 & 0 & 0 & 0 & 0\\
      0 & 0 & 1 & 0 & 0 & 0
      \end{array}\right).
  \end{align}
  From the differential equation~(\ref{eq_master_equation}), we get the time evolution for
   $(\boldsymbol{\Psi}_{\lambda_E,\lambda_X})_{n} (t)$
   \begin{align}
    \frac{(\boldsymbol{\Psi}_{\lambda_E,\lambda_X})_{n} (t)}{\partial t}&=\underset{m}{\sum}\biggl\{W_{nm}{\rm{e}}^{-\lambda_{E}
    \Delta E_{nm}-\lambda_{X}\Delta X_{nm}}-\delta_{nm}\Bigl[\underset{l(\neq n)}{\sum}W_{lm}\Bigr]\biggr\}(\boldsymbol{\Psi}_{\lambda_E,\lambda_X})_{m} (t)\\
    &=\underset{m}{\sum}(\boldsymbol{\sf{L}}_{\lambda_{E},\lambda_{X}})_{nm}(\boldsymbol{\Psi}_{\lambda_E,\lambda_X})_{m} (t).
   \end{align}
   This way we get the evolution matrix $\boldsymbol{\sf{L}}_{\lambda_{E},\lambda_{X}}$(\ref{eq_L_lambda}) .
  
\end{widetext}

\bibliography{paper5}

\end{document}